\documentclass[showpacs,showkeys,preprintnumbers, amsmath,amssymb,preprint]{revtex4-1}
\usepackage{graphicx}
\usepackage{epsfig}
\usepackage{bm}

\usepackage{fontenc}
\usepackage{graphicx}
\usepackage[dvips]{hyperref}
\usepackage{bm}
\usepackage{amsmath}

\usepackage{color}
\usepackage{soul}

\begin{document}

\title{Greybody Factors of Massive Charged Fermionic Fields in a Charged Two-Dimensional Dilatonic Black Hole}
\author{Ram\'{o}n B\'{e}car}
\email{rbecar@uct.cl}
\affiliation{Departamento de Ciencias Matem\'{a}ticas y F\'{\i}sicas, Universidad Cat\'{o}%
lica de Temuco, Montt 56, Casilla 15-D, Temuco, Chile}
\author{P. A. Gonz\'{a}lez}
\email{pablo.gonzalez@udp.cl}
\affiliation{Facultad de Ingenier\'{\i}a, Universidad Diego Portales, Avenida Ej\'{e}%
rcito Libertador 441, Casilla 298-V, Santiago, Chile.}
\author{Joel Saavedra}
\email{joel.saavedra@ucv.cl} \affiliation{Instituto de
F\'{i}sica, Pontificia Universidad Cat\'olica de Valpara\'{i}so,
Casilla 4950, Valpara\'{i}so, Chile.}
\author{Yerko V\'{a}squez.}
\email{yvasquez@userena.cl}
\affiliation{Departamento de F\'{\i}sica, Facultad de Ciencias, Universidad de La Serena,\\
Avenida Cisternas 1200, La Serena, Chile.}
\date{\today }

\begin{abstract}
We study massive charged fermionic perturbations in the background of a charged two-dimensional dilatonic black hole, and we solve the Dirac equation analytically. Then, we compute the reflection and transmission coefficients and the absorption cross section for massive charged fermionic fields, and we show that the absorption cross section vanishes at the low and high frequency limits. However, there is a range of frequencies where the absorption cross section is not null. Furthermore, we study the effect of the mass and electric charge of the fermionic field over the absorption cross section.
\end{abstract}
\maketitle
\section{Introduction}

In order to find a clue on the Quantum Gravity problem in  spacetime for which $D=4$, a very rich model of different lower dimensional gravity  has been developed. In the particular  case of $D=2$, it is well known that  the Einstein-Hilbert action has been used as the gravity sector. However, this model 
is  locally trivial because the Einstein-Hilbert action in $D=2$ is just a topological invariant (Gauss-Bonnet theorem). If we want to obtain the dynamical degree of freedom, we need to couple this action with different fields besides the gravitational one. Under this perspective, the dilatonic field has shown a very rich structure and includes black hole solutions. The dilatonic field naturally arises, for instance, in the compactifications from higher dimensional gravity  or from  string theory. Two dimensional dilatonic-gravity has black hole solutions that play an important role and reveal various physical aspects such as   spacetime geometry,  the quantization of gravity, and also the physics related to string theory \cite{Witten:1991yr, Teo:1998kp, McGuigan:1991qp}.  Furthermore, technical simplifications in two dimensions often lead to exact results, and it is hoped that this might help to address some of the conceptual problems posed by quantum gravity in higher dimensions. The exact solvability of two-dimensional models of gravity have been a useful tool for research in black hole thermodynamics \cite{Lemos:1996bq,Youm:1999xn, Davis:2004xb, Grumiller:2007ju, Quevedo:2009ei, Belhaj:2013vza}. Such lines of research are provided to give deeper understanding of some key issues, including the microscopic origin of black hole entropy \cite{Myers:1994sg, Sadeghi:2007kn, Hyun:2007ii}, and the final stages of black hole evaporation  \cite{Kim:1999ig, Vagenas:2001sm, Easson:2002tg}. For a review of two dimensional dilaton gravity see \cite{Grumiller:2002nm}, in the specific subject of black hole physics, these are  several studies that have contributed to understanding the scattering and absorption properties of waves in  black holes. Because spacetime geometry  surrounding a black hole is non-trivial, the Hawking radiation emitted at the event horizon is modified by this geometry, and therefore  an asymptotic observer measuring the black hole thermal spectrum,   will  measure a modified spectrum and no longer the well known black body thermal spectrum \cite{Maldacena:1996ix}.  The factors that modify the emitted spectrum  of black holes
are known as greybody factors and can be obtained through the classical scattering for fields under the influence of a black hole. Because the nature of Hawking radiation is quantum,   the study of greybody factors allows  increases the semiclassical gravity dictionary, and also  gives further steps into the quantum nature of black holes,  for a review about this topic see \cite{Harmark:2007jy}.

In the present work we study the reflection and  transmission coefficients, and the greybody factors 
of massive charged fermions fields on the background of two-dimensional charged dilatonic black holes \cite{Witten:1991yr}, \cite{Frolov:2000jh}. Greybody factors for scalar and fermionic field perturbations on the background of black holes have received great attention. In this context, it was shown that for all spherically symmetric black holes, the low energy cross section for massless minimally-coupled scalar fields is always the area of the horizon, where the contribution to the absorption cross section comes from the mode with lowest angular momentum \cite{StarobinskyII, Starobinsky, Das:1996we}. However, for asymptotically AdS and Lifshitz black holes, it was observed
that, at the low frequency limit there is a range of modes with highest angular momentum, which contribute to the absorption cross section in addition to the mode with lowest angular momentum \cite{Gonzalez:2010ht, Gonzalez:2010vv, Gonzalez:2011du,Gonzalez:2012xc}. Also, it was observed that the absorption cross section for the three dimensional warped AdS black hole is larger than the area, even if the $s$-wave limit is considered \cite{Oh:2009if}. Recently it has been found that
the  zero-angular-momentum greybody factors for non-minimally coupled scalar fields in four-dimensional Schwarzschild-de Sitter spacetime tends to zero around the zero-frequency limit \cite{Crispino:2013pya}. Otherwise, for fermionic fields, it was shown that the absorption probability for bulk massive Dirac fermions in higher-dimensional Schwarzschild black hole
increases with the dimensionality of the spacetime and
decreases as the angular momentum increases. For this spacetime, it was also revealed that the absorption probability depended on the mass of the emitted field, that is, that the absorption probability decreases or increases depending on the range of energy when the mass of the field increases. Also, it has been observed that the absorption probability increases for higher radii of the event horizon \cite{Rogatko:2009jp}, see for instance \cite{Moderski:2008nq, Gibbons:2008gg} for the decay of Dirac fields in higher dimensional black holes.
For further reference, massive charged scalar field perturbations of the Kerr-Newman black hole background were studied in \cite{Konoplya:2013rxa, Konoplya:2014sna},  the absorption of photons and fermions by black holes in four-dimensions in \cite{Gubser:1997cm}, the fermion absorption cross section of a Schwarzschild black hole in \cite{Doran:2005vm},  and charged fermionic perturbations in the Reissner-Nordstrom anti-de Sitter black hole background  in \cite{Cai:2010tr}. For higher-dimensional black hole background see \cite{Jung:2004nh, Jung:2005sw}. Furthermore, fermionic perturbations on the background of two-dimensional dilatonic black holes have been studied in which it was shown that the absorption cross section vanishes at the low and high frequency limits. However, there is a range of frequencies where the absorption cross section is not null \cite{Becar:2014aka}. Besides, charged fermionic field perturbations have been studied in order to obtain the quasinormal modes and to study the stability of these black holes \cite{Becar:2014jia}.
As such this paper is organized as follows. In Sec. \ref{FP}, we study massive charged fermionic perturbations in the background of two-dimensional dilatonic black holes, and in Sec. \ref{coeff} we calculate the reflection and the transmission coefficients, and the absorption cross section. Finally, our conclusions are in Sec. \ref{remarks}.

\section{Massive Charged fermionic perturbations in  two-dimensional charged dilatonic black holes}

\label{FP}
Let us begin with the effective action of Maxwell-gravity coupled to a
dilatonic field $\phi$  \cite{McGuigan:1991qp}
\begin{equation}
S=\frac{1}{2\pi }\int d^{2}x\sqrt{-g}e^{-2\phi }\left( R-4(\nabla \phi
)^{2}-\lambda -\frac{1}{4}F_{\mu \nu }F^{\mu \nu }\right)~,  \label{accion}
\end{equation}%
where $R$ is the Ricci scalar, $\lambda$ is the central charge, and $F_{\mu\nu}$ is the electromagnetic strength tensor.
If we perform the variation of the metric, gauge, and dilaton field, we obtain the following equations of motions.
\begin{eqnarray}
\nonumber R_{\mu \nu }-2\nabla _{\mu }\nabla _{\nu }\phi -\frac{1}{2}F_{\mu \sigma
}F_{\nu }^{\sigma } &=&0~,  \label{betag} \\
\nonumber \nabla _{\nu }\left( e^{-2\phi }F^{\mu \nu }\right)  &=&0\text{ }, \\
R-4\nabla _{\mu }\nabla ^{\mu }\phi +4\nabla _{\mu }\phi \nabla ^{\mu }\phi
-\lambda -\frac{1}{4}F_{\mu \nu }F^{\mu \nu } &=&0~.  \label{betaphi}
\end{eqnarray}%
In order to describe the black hole solution, we considered the following form of the static metric for charged black holes
\begin{equation}
ds^{2}=-f(r)dt ^{2}+\frac{dr^{2}}{f(r)}~,  \label{metrica1}
\end{equation}%
in this expression, $f(r)=1-2me^{-Qr}+q^{2}e^{-2Qr}$, $\phi =\phi _{0}-\frac{Q}{2}r$, and $%
F_{tr}=\sqrt{2}Qqe^{-Qr}$.  We used that $\lambda=-Q^2$  because the asymptotic flatness condition for the spacetime require. It is well known that  $m$ and $q$ (free parameters) are proportional to the black hole mass and  charge, respectively. The positions of the horizons are given by
\begin{equation}
r_{\pm }=\frac{1}{Q}\ln \left( m\pm \sqrt{m^{2}-q^{2}}\right) \text{ },
\end{equation}
we can obtain one single horizon solution ($r_{+}$) if the following condition is fulfilled
 $m^{2}-q^{2}\geqslant 0$, from which it is straightforward to see that $m^{2}=q^{2}$ corresponds an extremal case, where $r_{+}=r_{-}$. On the other hand, using the coordinate transformation $y=e^{-Qr}$ yields $f(y)=1-2my+q^{2}y^{2}$, where the spatial infinity is now located at $y=0$. We can see, this solution represents a
well-known string-theoretic black hole \cite%
{Teo:1998kp,McGuigan:1991qp,Witten:1991yr}. As it is well known, charged fermionic
perturbations on the background of two-dimensional charged dilatonic black hole are governed by the Dirac equation
\begin{equation}
\left( \gamma ^{\mu }\left( \nabla _{\mu }+iq^{\prime }A_{\mu }\right)
+m^{\prime}\right) \psi =0~,  \label{DE}
\end{equation}%
where $A_{\mu}$ denotes the electromagnetic potential, $q^{\prime }$ and $m^{\prime}$ denote the charge and the mass of the fermionic field $\psi$ respectively,
and 
\begin{equation}
\nabla _{\mu }=\partial _{\mu }+\frac{1}{2}\omega _{\text{ \ \ \ }\mu
}^{ab}J_{ab}~,
\end{equation}%
represents the covariant derivative $\nabla _{\mu }$. In this last expression  $J_{ab}=\frac{1}{4}\left[ \gamma _{a},\gamma _{b}\right]$ are the generators of the Lorentz group, and 
$\gamma ^{\mu }$ are the gamma matrices in curved spacetime. These are defined by
$\gamma ^{\mu }=e_{\text{ \ }a}^{\mu }\gamma ^{a}$,
where $\gamma ^{a}$ are the gamma matrices in flat spacetime. Here, we consider the following representation for the $2\times 2$ gamma matrices
\begin{equation}
\gamma ^{0}=i\sigma ^{2}~,\text{ \ }\gamma ^{1}=\sigma ^{1}~,
\end{equation}%
where $\sigma ^{i}$ are the Pauli matrices.
Now, in order to find the solution to the Dirac equation in this background we use the diagonal vielbein given by
\begin{equation}
e^{0}=\sqrt{f\left( r\right) }dt~,\text{ \ }e^{1}=\frac{1}{\sqrt{f\left(
r\right) }}dr~,
\end{equation}%
and from the null torsion condition
$de^{a}+\omega _{\text{ \ }b}^{a}\wedge e^{b}=0$~,
we obtain the spin connection
\begin{equation}
\omega ^{01}=\frac{f^{\prime }\left( r\right) }{2\sqrt{f\left( r\right) }}%
e^{0}.
\end{equation}%
Therefore, choosing the following ansatz for the fermionic field
\begin{equation}
\psi =\frac{1}{f\left( r\right) ^{1/4}}e^{-i\omega t}\left(
\begin{array}{c}
\psi _{1} \\
\psi _{2}%
\end{array}%
\right) ~,
\end{equation}%
we obtain the following coupled system of equations
\begin{eqnarray}
\sqrt{f}\partial _{r}\psi _{1}+\frac{i\omega }{\sqrt{f}}\psi _{1}-\frac{%
\sqrt{2}iqq^{\prime }}{\sqrt{f}}e^{-Qr}\psi _{1}+m^{\prime}\psi _{2} &=&0~  \notag
\label{system} \\
\sqrt{f}\partial _{r}\psi _{2}-\frac{i\omega }{\sqrt{f}}\psi _{2}+\frac{%
\sqrt{2}iqq^{\prime }}{\sqrt{f}}e^{-Qr}\psi _{2}+m^{\prime}\psi _{1} &=&0~.
\label{system}
\end{eqnarray}
Now, decoupling the above equations we obtain the following equation for $\psi _{1}$
\begin{gather}
\nonumber 2f\left( r\right) ^{2}\psi _{1}^{\prime \prime }(r)+f\left( r\right)
f^{\prime }\left( r\right) \psi _{1}^{\prime }(r)+e^{-2Qr}(4q^{2}q^{\prime
2}-4\sqrt{2}e^{Qr}qq^{\prime }\omega +2e^{2Qr}\omega ^{2}   \\
-2e^{Qr}(m^{\prime 2}e^{Qr}-\sqrt{2}iqq^{\prime }Q)f(r)-ie^{Qr}\left( -\sqrt{2}%
qq^{\prime }+e^{Qr}\omega \right) f^{\prime }(r))\psi _{1}(r)=0~,
\label{radial}
\end{gather}
and now performing the transformation $y=e^{-Qr}$, Eq. (\ref%
{radial}) becomes
\begin{equation}
\psi _{1}^{\prime \prime }\left( y\right) +\left( \frac{1}{y}+\frac{1/2}{%
y-y_{+}}+\frac{1/2}{y-y_{-}}\right) \psi _{1}^{\prime }\left( y\right)
+\left( \frac{A_{1}}{y}+\frac{A_{2}}{y-y_{+}}+\frac{A_{3}}{y-y_{-}}\right)
\frac{1}{y\left( y-y_{+}\right) \left( y-y_{-}\right) }\psi _{1}\left(
y\right) =0~,  \label{diff}
\end{equation}
where $y_{\pm }$ are the roots of the function $f(y)=1-2my+q^{2}y^{2}$, which are
given by
\begin{equation}
y_{\pm }=\frac{m\mp \sqrt{m^{2}-q^{2}}}{q^{2}}\text{ },
\end{equation}
and the constants $A_{1}$, $A_{2}$ and $A_{3}$ are defined by the
expressions:
\begin{eqnarray}
A_{1} &=&\frac{1}{q^{2}Q^{2}}\left( \omega ^{2}-m^{\prime 2}\right) \text{ }, \\
A_{2} &=&y_{+}\left( y_{+}-y_{-}\right) \left( \frac{1}{16}-\left( \frac{1}{4%
}-\frac{i\omega }{q^{2}Qy_{+}\left( y_{+}-y_{-}\right) }+\frac{\sqrt{2}%
iq^{\prime }}{qQ\left( y_{+}-y_{-}\right) }\right) ^{2}\right) \text{ }, \\
A_{3} &=&-y_{-}\left( y_{+}-y_{-}\right) \left( \frac{1}{16}-\left( \frac{1}{%
4}+\frac{i\omega }{q^{2}Qy_{-}\left( y_{+}-y_{-}\right) }-\frac{\sqrt{2}%
iq^{\prime }}{qQ\left( y_{+}-y_{-}\right) }\right) ^{2}\right) \text{ }.
\end{eqnarray}
Additionally, we perform the change of variable $z=\left(\frac{y_{-}}{y_{+}}\right)\left(\frac{y-y_{+}}{y-y_{-}}\right)$, and making the substitution
\begin{equation}
\psi _{1}(z)=z^{\alpha }\left( 1-z\right) ^{\beta }F\left( z\right)~, 
\end{equation}
in Eq. (\ref{diff}), we obtain the following equation for $F\left(z\right)$
\begin{equation}\label{HE}
z\left( 1-z\right) F^{\prime \prime }\left( z\right) +\left( c-\left(
1+a+b\right) z\right) F^{\prime }\left( z\right) -abF\left( z\right) =0~, 
\end{equation}
where
\begin{equation}
\alpha_{\pm}=\frac{1}{4}\pm\left( \frac{1}{4}-\frac{i\sqrt{2}q^{\prime}}{Qq(y_{-}-y_{+})}+\frac{i\omega y_{-}}{Q(y_{-}-y_{+})}\right)~, 
\end{equation}
\begin{equation}
\beta_{\pm} =\pm \frac{i \sqrt{\omega ^2-m^{\prime 2}}}{Q}~.
\end{equation}
Therefore, as Eq. (\ref{HE}) corresponds to the hypergeometric equation, its solution is given by
\begin{equation}
\psi _{1}=C_{1}z^{\alpha }\left( 1-z\right) ^{\beta }{_2F_1}\left(
a,b,c,z\right) +C_{2}z^{1/2-\alpha }\left( 1-z\right) ^{\beta }{_2F_1}\left(
a-c+1,b-c+1,2-c,z\right)~, 
\end{equation}
which has three regular singular points at $z=0$, $z=1$ and $z = \infty$. Here, $_2F_1(a,b,c;z)$ denotes the Gauss hypergeometric function and $C_1$, $C_2$ are integration constants and
\begin{equation}
a=\frac{1}{2}+2\alpha +\beta+\frac{i\omega}{Q}~, 
\end{equation}
\begin{equation}
b=\beta-\frac{i\omega}{Q}~, 
\end{equation}
\begin{equation}
c=\frac{1}{2}+2\alpha~. 
\end{equation}
Now, imposing boundary conditions at the horizon, i.e., that there is only ingoing waves, and choosing $\alpha=\alpha_{-}$ implies that $C_{2}=0$. Thus,  the solution for $\psi_{1}$reduces to
\begin{equation}
\label{Rhorizon}
\psi _{1}=C_1z^{\alpha }\left( 1-z\right) ^{\beta }{_2F_1}\left(
a,b,c,z\right)~.
\end{equation}

Otherwise, in order to find the solution for $\psi_{2}$, we use the change of variable defined before, i.e.,  $z=\left(\frac{y_{-}}{y_{+}}\right)\left(\frac{y-y_{+}}{y-y_{-}}\right)$. Thus, the second equation of the system (\ref{system}), can be written as
\begin{equation}\label{psi2}
\psi_{2}^{\prime}(z)-\frac{i\omega (y_{-}-y_{+})}{(1-z)(y_{-}-y_{+}z) Q f(z)}\psi_{2}+\frac{\sqrt{2}iqq^{\prime}y_{+}y_{-}(y_{-}-y_{+})}{(y_{-}-y_{+}z)^2Qf(z)}\psi_2+\frac{m^{\prime}(y_{-}-y_{+})}{(1-z)(y_{-}-y_{+}z)\sqrt{f(z)}Q}\psi_1=0~. 
\end{equation}
Now, by using the integrating factor $I$ given by:
\begin{equation}
I=z^{-\frac{i}{2Q\sqrt{m^2-q^2}}\left( -\sqrt{2}qq^{\prime}+(m+\sqrt{m^2-q^2}) \omega\right)}(1-z)^{\frac{i\omega}{Q}}=z^{\alpha}(1-z)^{\frac{i\omega}{Q}}~,
\end{equation}
we integrate Eq. (\ref{psi2}) and we obtain the solution
\begin{equation}
\psi _{2}=-\frac{C_1m^{\prime}}{Qz^{\alpha}(1-z)^{\frac{i\omega}{Q}}}\int z^{\prime c-1}\left(
1-z^{\prime }\right) ^{a-c-1}{_2F_1}\left( a,b,c,z^{\prime }\right)
dz^{\prime }~,
\end{equation}
which can be written as
\begin{equation}\label{psi2horizon}
\psi _{2}=-\frac{C_1m^{\prime}z^{\frac{1}{2}+\alpha}(1-z)^{\frac{i\sqrt{\omega^2-m^{\prime 2}}}{Q}}}{Q(\frac{1}{2}+2\alpha)} {}_2F_1\left( a,b+1,c+1,z\right)~,
\end{equation}
by using the relation
\begin{equation}
\int z^{c-1}\left( 1-z\right) ^{a-c-1}{_2F_1}\left( a,b,c,z\right) dz=\left(
1-z\right) ^{a-c}z^{c}\frac{{_2F_1}\left( a,b+1,c+1,z\right) }{c}~.
\end{equation}

\section{Reflection coefficient, transmission coefficient, and absorption cross section}
\label{coeff}
The reflection and transmission coefficients depend on the behavior
of the radial function, at the horizon and at the asymptotic
infinity, and  are defined by
\begin{equation}\label{reflectiond}\
\mathcal{R} :=\left|\frac{\mathcal{F}_{\mbox{\tiny asymp}}^{\mbox{\tiny
out}}}{\mathcal{F}_{\mbox{\tiny asymp}}^{\mbox{\tiny in}}}\right|; \; \mathcal{T}:=\left|\frac{\mathcal{F}_{\mbox{\tiny
hor}}^{\mbox{\tiny in}}}{\mathcal{F}_{\mbox{\tiny asymp}}^{\mbox{\tiny
in}}}\right|~,
\end{equation}
where $\mathcal{F}$ is the flux, and is given by 
\begin{equation}\label{flux}
\mathcal{F} =\sqrt{-g}\bar{\psi}\gamma ^{r}\psi~, 
\end{equation}
where, $\gamma ^{r}=e_{1}^{r}\gamma ^{1}$, $\bar{\psi}=\psi ^{\dagger
}\gamma ^{0}$,
$\sqrt{-g}=1$, 
and
$e_{1}^{r}=\sqrt{f\left(r\right)}$,
which yields
\begin{equation}\label{flux} 
\mathcal{F}=\left\vert \psi _{1}\right\vert ^{2}-\left\vert \psi _{2}\right\vert
^{2}~.
\end{equation}
The behavior of the fermionic field $\psi$ at the horizon is given by Eq.~(\ref{Rhorizon}) for $\psi_1$ and Eq.~(\ref{psi2horizon}) for $\psi_2$ in the limit $z\rightarrow 0$. Then, using Eq.~(\ref{flux}), we get the
flux at the horizon
\begin{equation}
\mathcal{F}
_{hor}^{in}= |C_{1}|^{2}~.
\end{equation}
Besides, in order to obtain the asymptotic behavior of $\psi_1$ and $\psi_2$ we use the Kummer's formula \cite{M. Abramowitz}:
\begin{eqnarray*}\label{KE}
{_2F_1}\left( a,b,c,z\right)  &=&\frac{\Gamma \left( c\right) \Gamma \left(
c-a-b\right) }{\Gamma \left( c-a\right) \Gamma \left( c-b\right) }%
{_2F_1}\left( a,b,a+b-c,1-z\right) + \\
&&\left( 1-z\right) ^{c-a-b}\frac{\Gamma \left( c\right) \Gamma \left(
a+b-c\right) }{\Gamma \left( a\right) \Gamma \left( b\right) }{_2F_1}\left(
c-a,c-b,c-a-b+1,1-z\right)~, 
\end{eqnarray*}
in Eq. (\ref{Rhorizon}) and Eq. (\ref{psi2horizon}), and by using Eq. (\ref{flux}) we obtain the flux at the asymptotic region $z\rightarrow 1$
\begin{equation}\label{fluxdinfinity}\
F_{asymp}= \left|A_1\right|^2+\left|A_2\right|^2-\left|B_1\right|^2-\left|B_2\right|^2~.
\end{equation}
where,
\begin{eqnarray*}
A_1  &=&C_1\frac{\Gamma \left( c\right) \Gamma \left(
c-a-b\right) }{\Gamma \left( c-a\right) \Gamma \left( c-b\right) }~,\\
A_2  &=&C_1\frac{\Gamma \left( c\right) \Gamma \left(a+b-c\right) }{\Gamma \left(a\right) \Gamma \left(b\right) }~,\\
B_1  &=&-\frac{C_1m^{\prime}}{Q}\frac{\Gamma \left( c\right) \Gamma \left(c-a-b\right) }{\Gamma \left(c+1-a\right) \Gamma \left(c-b\right) }~,\\
B_2  &=&-\frac{C_1m^{\prime}}{Q}\frac{\Gamma \left( c\right) \Gamma \left(a+b-c\right) }{\Gamma \left(a\right) \Gamma \left(b+1\right) }~,
\end{eqnarray*}
where in the last two equations we have used the property $\Gamma(c+1)=c \Gamma(c)$. Therefore, the reflection and transmission coefficients are given by
\begin{equation}
\mathcal{R}=\frac{|B_1|^{2}+|B_2|^{2}}{|A_1|^{2}+|A_2|^{2}}
~,\label{coef1}
\end{equation}
\begin{equation}
\mathcal{T}=\frac{|C_1|^{2}}{|A_1|^{2}+|A_2|^{2}}
~,\label{coef2}
\end{equation}
and the absorption cross section $\sigma_{abs}$, reads
\begin{equation}\label{absorptioncrosssection}\
\sigma_{abs}=\frac{1}{\omega}\frac{|C_1|^{2}}{|A_1|^{2}+|A_2|^{2}}~.
\end{equation}

\begin{figure}[h]
\begin{center}
\includegraphics[width=0.45\textwidth]{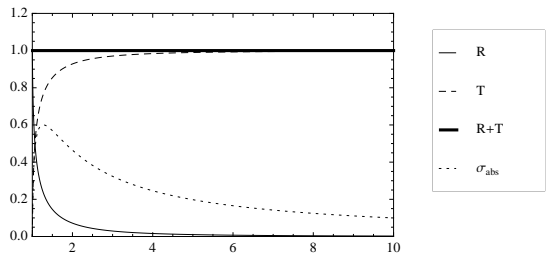}
\caption{The reflection coefficient $R$ (solid curve), the transmission coefficient $T$ (dashed curve),  $R+T$ (thick curve), and the absorption cross section $\sigma_{abs}$ (dotted curve) as a function of $\omega$, $(1\leq\omega)$; for $q=0.5$, $q^{\prime}=1$, $m=1$, $m^{\prime}=1$, and $Q=1$.}
\label{Coefficients}
\end{center}
\end{figure}

\begin{figure}[h]
\begin{center}
\includegraphics[width=0.45\textwidth]{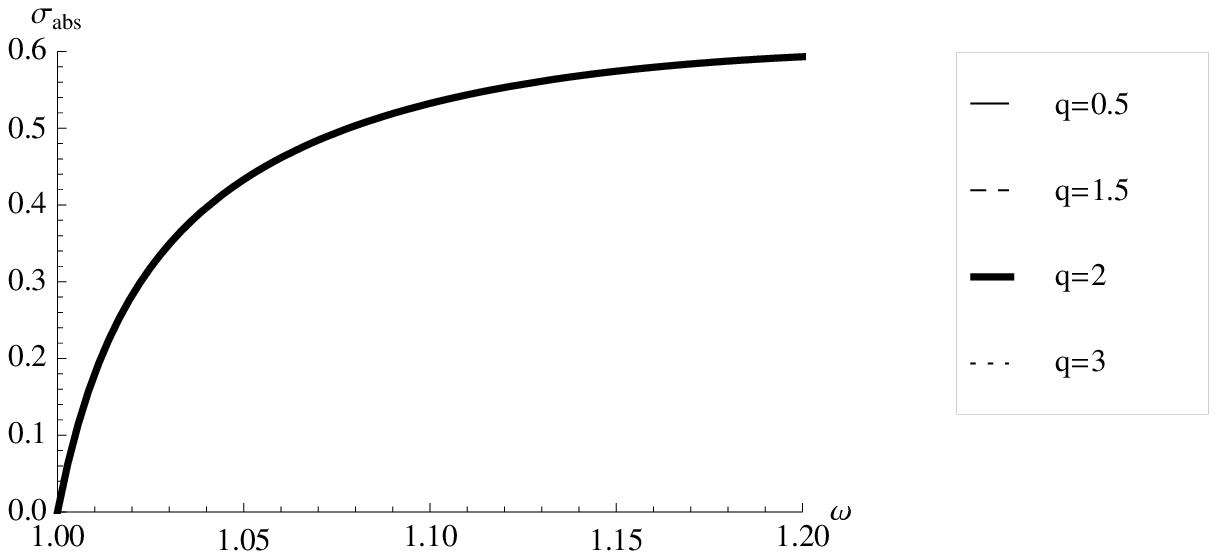}
\includegraphics[width=0.45\textwidth]{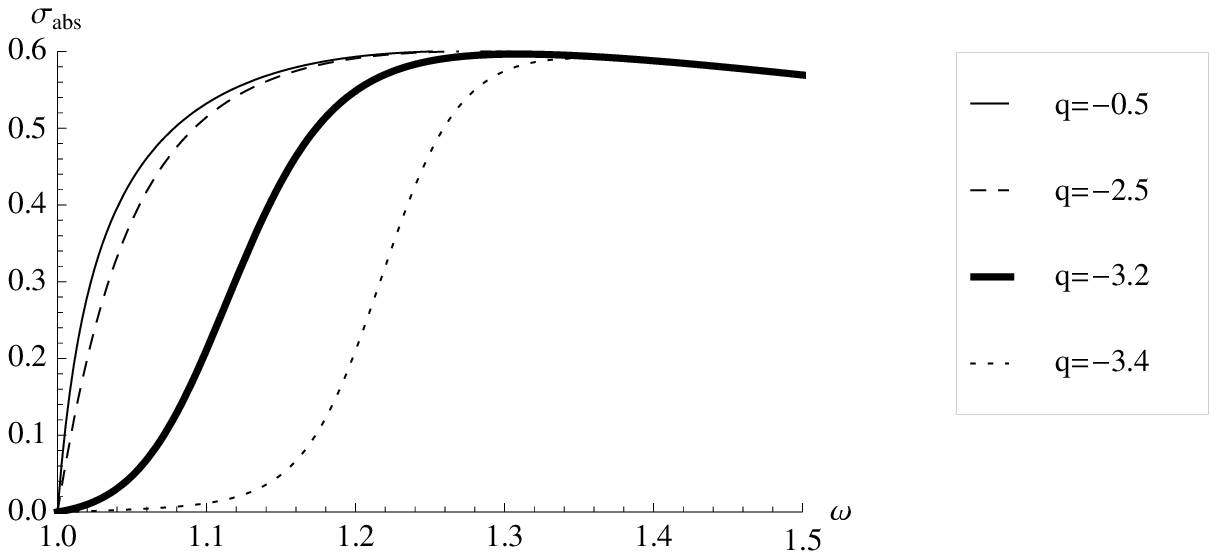}
\caption{The absorption cross section $\sigma_{abs}$ (dotted curve) as a function of $\omega$, $(1\leq\omega)$; for $q^{\prime}=-1$, $m=3.5$, $m^{\prime}=1$, $Q=1$, and $q=0.5, 1.5, 2, 3$ for the left figure, and $q=-0.5, -2.5,- 3.2, -3.4$ for the right figure.}
\label{Aq}
\end{center}
\end{figure}

\begin{figure}[h]
\begin{center}
\includegraphics[width=0.45\textwidth]{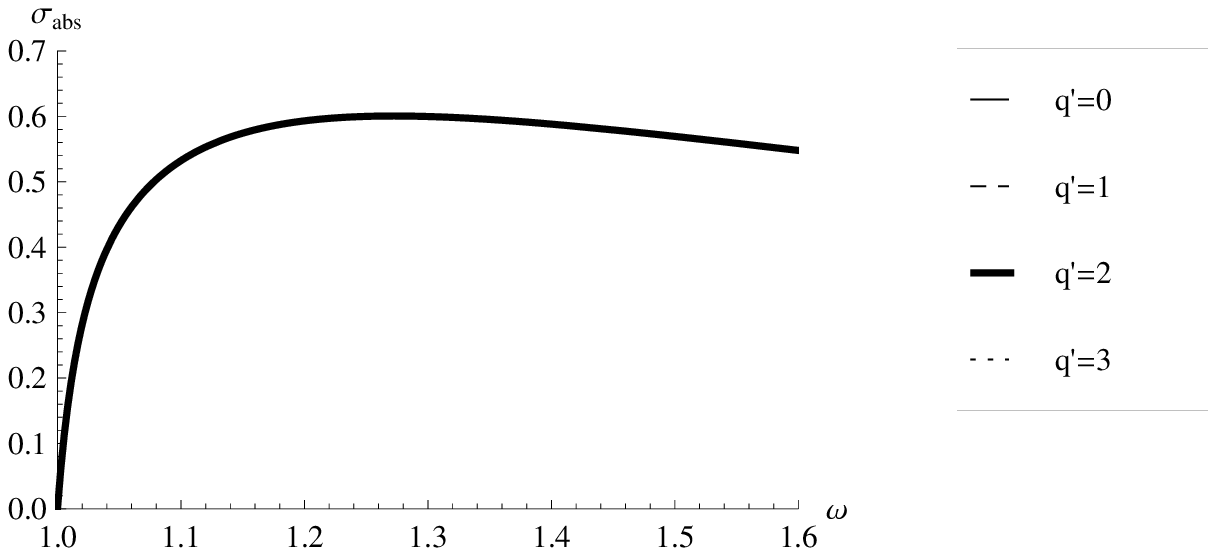}
\includegraphics[width=0.45\textwidth]{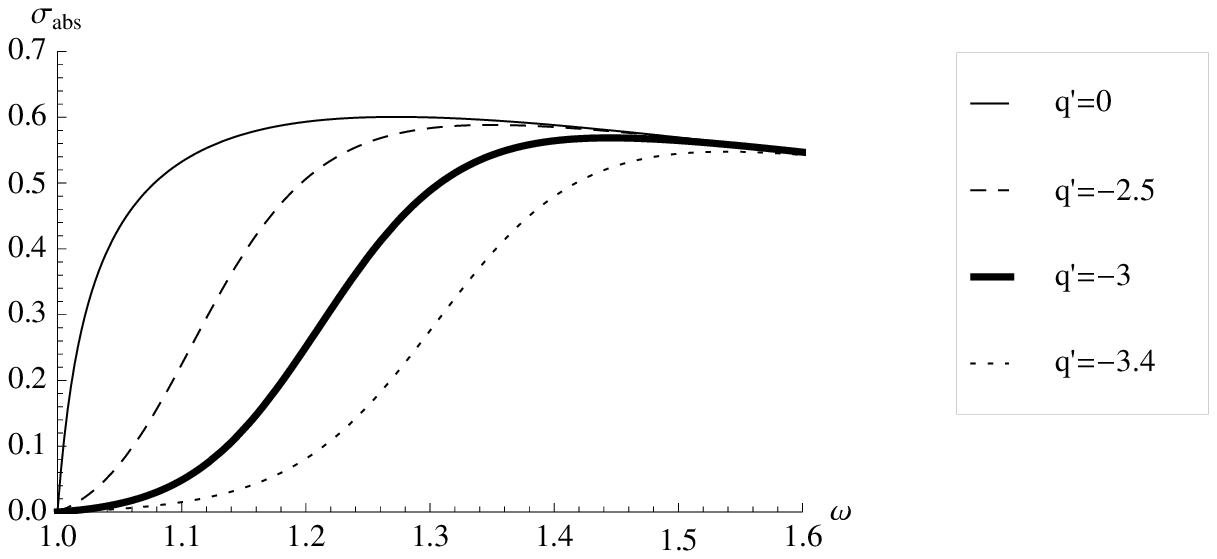}
\caption{The absorption cross section $\sigma_{abs}$ (dotted curve) as a function of $\omega$, $(1\leq\omega)$; for $m=3.5$, $m^{\prime}=1$, $Q=1$, $q=-1.5$ and $q^{\prime}=0, 1, 2, 3$ for the left figure, and $q^{\prime}=0, -2.5, -3, -3.4$ for the right figure.}
\label{Aqq}
\end{center}
\end{figure}

\begin{figure}[h]
\begin{center}
\includegraphics[width=0.45\textwidth]{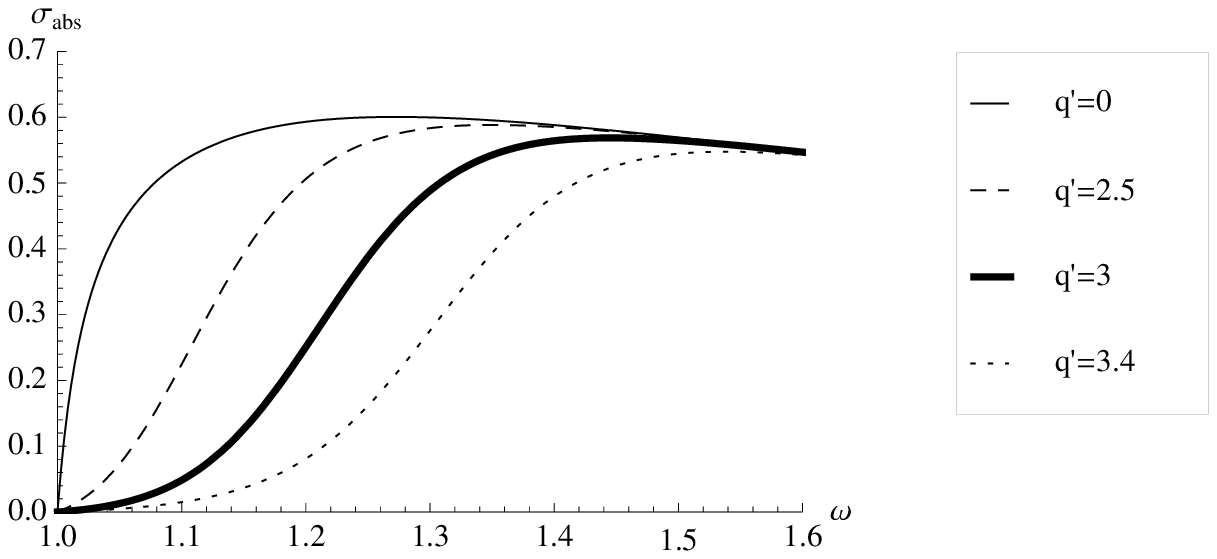}
\includegraphics[width=0.45\textwidth]{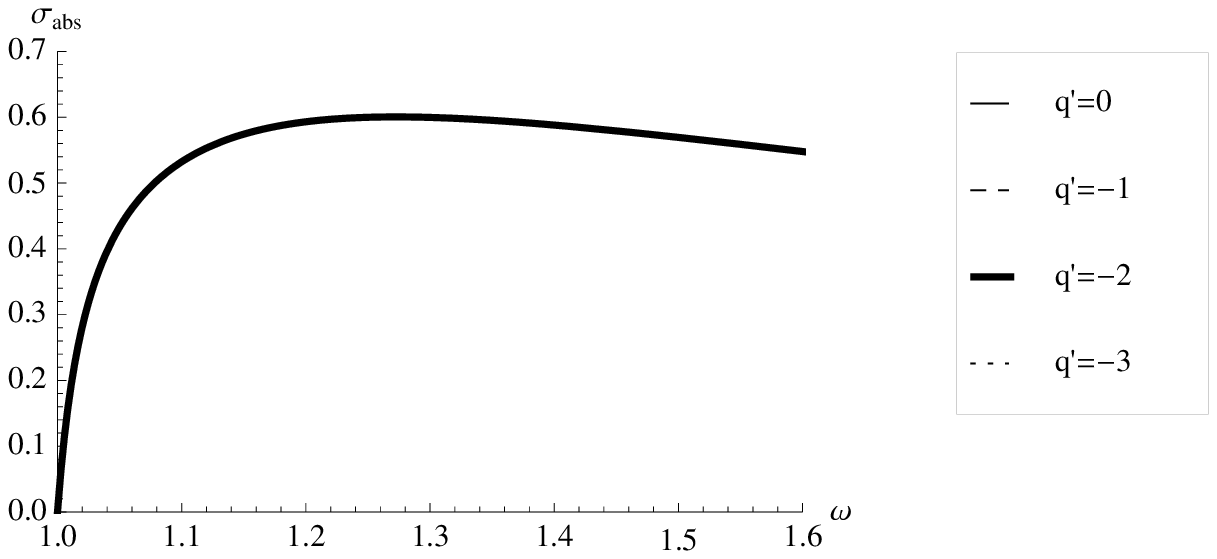}
\caption{The absorption cross section $\sigma_{abs}$ (dotted curve) as a function of $\omega$, $(1\leq\omega)$; for $q=1.5$, $m=3.5$, $m^{\prime}=1$, $Q=1$ and $q^{\prime}=0, 2.5, 3, 3.4$ for the left figure, and $q^{\prime}=0, -1, -2, -3$ for the right figure.}
\label{Aq1}
\end{center}
\end{figure}

\begin{figure}[h]
\begin{center}
\includegraphics[width=0.45\textwidth]{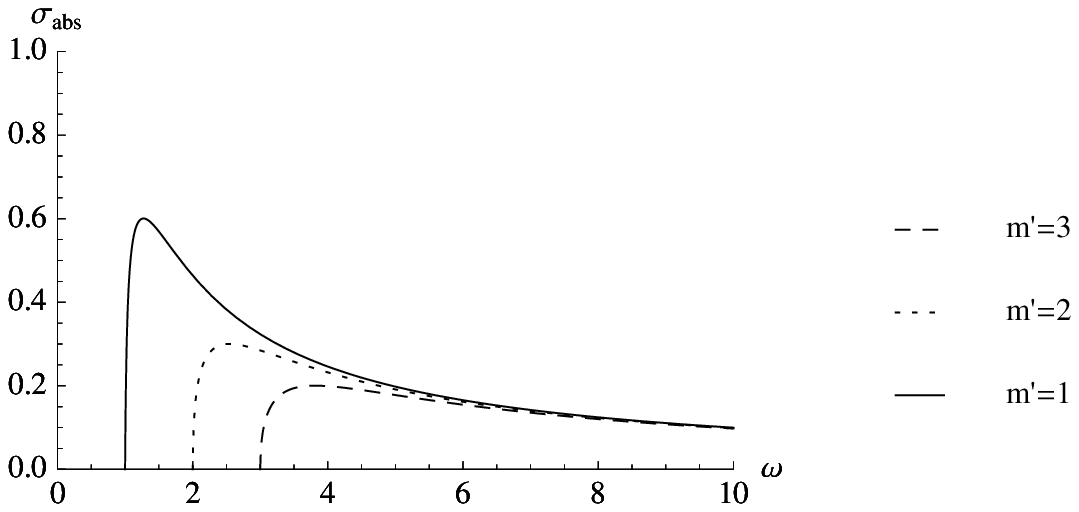}
\caption{The absorption cross section $\sigma_{abs}$ (dotted curve) as a function of $\omega$; for $q=0.5$, $q^{\prime}=1$, $m=1$, $Q=1$, and $m^{\prime}=1, 2, 3$.} 
\label{Amp}
\end{center}
\end{figure}

\begin{figure}[h]
\begin{center}
\includegraphics[width=0.45\textwidth]{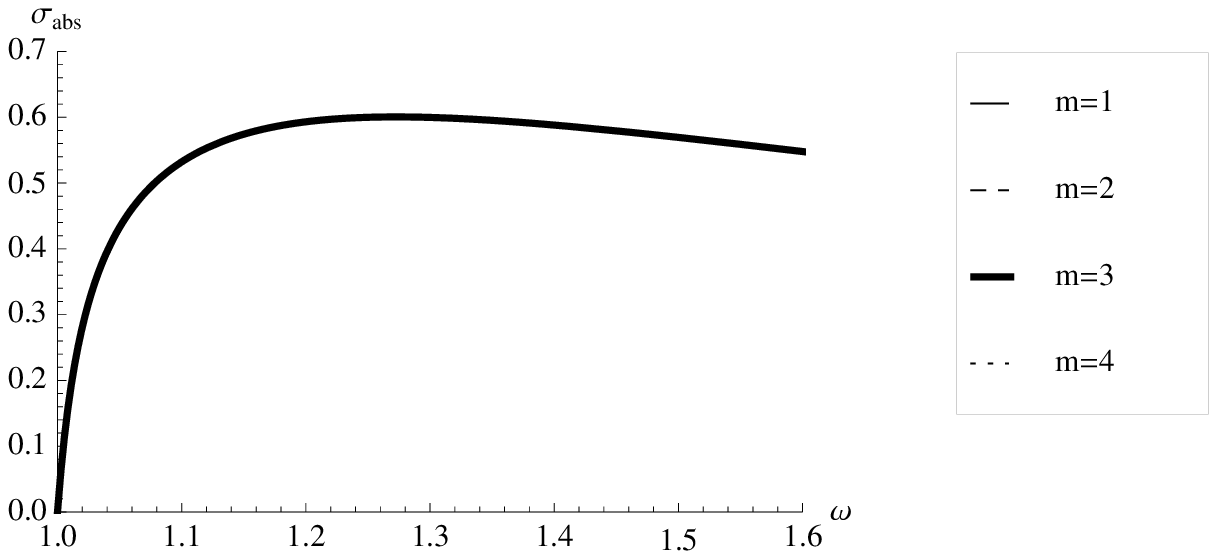}
\caption{The absorption cross section $\sigma_{abs}$ (dotted curve) as a function of $\omega$; for $q=0.5$, $q^{\prime}=1$, $m^{\prime}=1$, $Q=1$, and $m=1, 2, 3, 4$.} 
\label{Am}
\end{center}
\end{figure}

\begin{figure}[h]
\begin{center}
\includegraphics[width=0.45\textwidth]{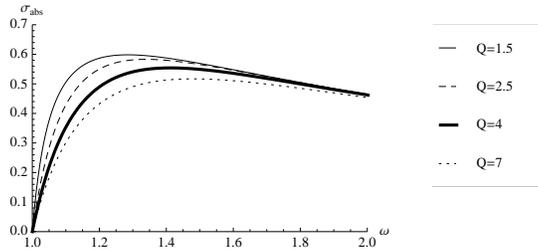}
\caption{The absorption cross section $\sigma_{abs}$ (dotted curve) as a function of $\omega$, $(1\leq\omega)$; for $q=0.5$, $q^{\prime}=1$, $m=1$, $m^{\prime}=1$, and $Q=1.5, 2.5, 4, 7$.}
\label{Aqm}
\end{center}
\end{figure}

Now, we perform a  numerical analysis of the reflection coefficient~(\ref{coef1}), transmission coefficient~(\ref{coef2}), and absorption cross section~(\ref{absorptioncrosssection}) of  two-dimensional charged dilatonic black holes, for charged fermionic fields.
In Fig.~(\ref{Coefficients}) we show the behavior of the reflection and transmission
coefficients and the absorption cross section, for charged fermionic fields for $q=0.5$, $q^{\prime}=1$, $m=1$, $m^{\prime}=1$, and $Q=1$.
Essentially, we found that the reflection coefficient is 1 at the low frequency limit, that is $\omega\approx m^{\prime}$, whereas for the high frequency limit this coefficient is null,  the opposite behavior of the transmission coefficient, with $\mathcal{R}+\mathcal{T}=1$.
Also, we observe that the absorption cross section is null at the low and high-frequency limits, but there is a range of frequencies for which the absorption cross section is not null, and also  has a maximum value.
In addition, In Fig.~(\ref{Aq}) we show the behavior of the absorption cross section for different (positive and negative) values of $q$, where we observe that the absorption cross section is null at the low and high-frequency limit, but there is a range of frequencies for which the absorption cross section is not null, and also  has a maximum value. Also, we observe in Figs.~(\ref{Aq}), (\ref{Aqq}) and (\ref{Aq1}) that for $qq^{\prime}>0$ the absorption cross section decreases when $qq^{\prime}$ increases, due to the electric repulsion. However, for $qq^{\prime}<0$ we found that the absorption cross section does not depend on the value of $qq^{\prime}$. 
Also, we observe in Fig.~(\ref{Amp}) that the absorption cross section increases if the mass of the fermionic field increases; however, beyond a certain value of the frequency, the absorption cross section is
constant and null for the high-frequency limit. On the other hand, in Fig.~(\ref{Am}), we plot the absorption cross section for different values of $m$ and we observe that it does not depend on the mass of the black hole. Finally, we observe that the absorption cross section increase if $Q$ decrease, see Fig.~(\ref{Aqm}).


\section{Conclusions}
\label{remarks}
In this work we have studied massive charged fermionic perturbations on the background of two-dimensional charged dilatonic black holes, and we have computed the reflection and transmission coefficients, and the absorption cross section, and we have shown numerically that the absorption cross section vanishes at the low and high frequency limits. Therefore, a wave emitted from the horizon, with low or high frequency, does not reach infinity and is totally reflected, since the fraction of particles  penetrating the potential barrier vanishes; however, we have shown that there is a range of frequencies where the absorption cross section is not null. The reflection coefficient is 1 at the low frequency limit and null for the high frequency limit,  demonstrating behavior opposite  of the transmission coefficient, with $\mathcal{R}+\mathcal{T}=1$.
It is worth  mentioning that these results, greybody factors, are consistent with other geometries of dilatonic black holes \cite{Kim:1995hy, Abedi:2013xua, Becar:2014aka}. Also, we have studied the effect of the electric charge of the fermionic field over the absorption cross section, and we have observed different behaviors depending on the sign and the value of the product of the charges $\propto qq^{\prime}$. That is, for $qq^{\prime}>0$ we have found that the absorption cross section decreases when $qq^{\prime}$ increases, due to the electric repulsion. However, for $qq^{\prime}<0$ we have found that the absorption cross section does not depend on the value of $qq^{\prime}$, and for this case we obtain the same value of the absorption cross section as for the case $q^{\prime}=0$. Also, we have found that the absorption cross section increases if the mass of the fermionic field increases; however, beyond a certain value of the frequency, the absorption cross section is
constant. Also, we have found that the absorption cross section for massive charged fermionic fields in a charged two-dimensional dilatonic black hole does not depend on the mass of the black hole. 

\section*{Acknowledgments}

This work was funded by Comisi\'{o}n
Nacional de Ciencias y Tecnolog\'{i}a through FONDECYT Grants
11140674 (PAG), 1110076 (JS) and 11121148 (YV) and by DI-PUCV Grant 123713
(JS). R. B. acknowledges the hospitality of the Universidad Diego Portales where part of this work was undertaken.

\end{document}